\documentclass[a4paper,12pt]{article}
\usepackage{amsfonts,slashed}
\usepackage{url}
\usepackage{latexsym}
\usepackage{amsfonts}
\usepackage{epsfig}
\usepackage{latexsym,amssymb}
\usepackage{amsmath,amssymb,amsthm}
\usepackage{ifpdf}
\usepackage{cite}
\usepackage[pdftex]{hyperref}
\usepackage{tikz}
\ifx\pdfoutput\undefined
   \pdffalse
\else
   \pdfoutput=1
   \pdftrue
\fi

\numberwithin{equation}{section}
\def\be{\begin{equation}}
\def\ee{\end{equation}}
\def\ba{\begin{array}}
\def\ea{\end{array}}

\newcommand{\bea}{\begin{eqnarray}}
\newcommand{\eea}{\end{eqnarray}}

\newcommand{\bbox}{\lower.2ex\hbox{$\Box$}}

\def\bfone{\relax{\rm 1\kern-.35em 1}}
\def\bfzero{\relax{\rm 0\kern -.45 em 0}}
\usetikzlibrary{trees}
\usetikzlibrary{positioning,shadows,arrows,decorations.pathmorphing,decorations.markings}
\tikzstyle{block}=[draw opacity=0.7,line width=1.4cm]

\begin{document}

\title{On the Hidden Maxwell Superalgebra underlying $D=4$ Supergravity}
\author{D. M. Pe\~{n}afiel$^{1,2,3}$\thanks{%
diegomolina@udec.cl}, L. Ravera$^{2,3}$\thanks{%
lucrezia.ravera@polito.it} \\
{\small $^{1}$\textit{Departamento de F\'{\i}sica, Universidad de Concepci\'{o}n,}} \\
{\small Casilla 160-C, Concepci\'{o}n, Chile}\\
{\small $^{2}$\textit{DISAT, Politecnico di Torino}}\\
{\small Corso Duca degli Abruzzi 24, I-10129 Torino, Italia}\\
{\small $^{3}$\textit{Istituto Nazionale di Fisica Nucleare (INFN)}}\\
{\small Sezione di Torino, Via Pietro Giuria 1, 10125, Torino, Italia}}
\maketitle

\vskip 1 cm

\begin{center}
{\small \textbf{Abstract} }
\end{center}

In this work, we expand the hidden $AdS$-Lorentz superalgebra underlying $D=4$ supergravity, reaching a (hidden) Maxwell superalgebra. The latter can be viewed as an extension involving cosmological constant of the superalgebra underlying $D=4$ supergravity in flat spacetime. We write the Maurer-Cartan equations in this context and we find some interesting extensions of the antisymmetric $3$-form $A^{(3)}$ appearing in the Free Differential Algebra in Minkowski space. The structure of Free Differential Algebras is obtained by considering the zero curvature equations. We write the parametrization of $A^{(3)}$ in terms of $1$-forms and we rend the topological features of its extensions manifest. We interestingly find out that the structure of these extensions, and consequently the structure of the corresponding boundary contribution $dA^{(3)}$, strongly depends on the form of the extra fermionic generator appearing in the hidden Maxwell superalgebra. The model we develop in this work is defined in an enlarged superspace with respect to the ordinary one, and the extra bosonic and fermionic $1$-forms required for the closure of the hidden Maxwell superalgebra must be considered as physical fields in this enlarged superspace.

\vskip 1 cm \eject
\numberwithin{equation}{section}

\section{Introduction}

Supergravity theories in various spacetime dimensions $4 \leq D \leq11$ have a field
content that generically includes the metric, the gravitino, a set of $1$-form gauge potentials, and $(p + 1)$-form gauge potentials of various $p\leq9$, and they are properly discussed in the context of Free Differential Algebras (FDAs) \footnote{The FDAs framework is analogous to the Cartan Integrable Systems (CIS) one.}. The FDAs are, in a sense to be described, cohomological extensions of normal Lie algebras for superalgebras, and their structure is obtained by considering the zero curvature equations.

In particular, in the framework of FDAs, the structure of $D=11$ supergravity, first constructed in \cite{Cremmer}, was then reconsidered in \cite{D'AuriaFre}, adopting the superspace geometric approach.
In the same paper, the supersymmetric FDA was also investigated in order to see whether the FDA formulation  could be interpreted in terms of an ordinary Lie superalgebra in its dual Maurer-Cartan formulation \footnote{The supergroup structure allows a deeper understanding of symmetries and topological properties of the theory.}.
This was proven to be true, and the existence of a hidden superalgebra underlying the theory was presented for the first time.
In fact, in \cite{D'AuriaFre}, the authors proved that the FDA underlying $D=11$ supergravity can be traded with a Lie superalgebra which contains, besides the Poincar\'{e} superalgebra, also new bosonic $1$-forms and a nilpotent fermionic generator $Q'$, necessary for the closure of the superalgebra.

Later, in \cite{Castellani}, the authors wonder whether eleven dimensional supergravity can be decontracted
into a non-abelian (gauged) model. This problem was reduced to that of finding an algebra whose contraction yields the $D=11$ algebra of \cite{D'AuriaFre}. In the same paper, they also considered the $D=4$ case, in order to introduce their approach through a toy-model; However, the four dimensional gauged case results of some interest, since its algebraic form (presented in \cite{Castellani}) corresponds to a ``hidden $AdS$-Lorentz-like superalgebra", an extension with an extra nilpotent fermionic generator of the $AdS$-Lorentz superalgebra presented and largely discussed in \cite{Gauss}. 

In particular, in \cite{Gauss} the authors explored the supersymmetry invariance of an extension of minimal $D=4$ supergravity in the presence of a non-trivial boundary, presenting the explicit construction of
the $\mathcal{N}=1$, $D=4$ $AdS$-Lorentz supergravity bulk Lagragian in the rheonomic framework. They developed a peculiar way to
introduce a generalized supersymmetric cosmological term to supergravity.
Then, by studying the supersymmetry invariance of the Lagrangian in the presence of a non-trivial boundary, they interestingly found that the supersymmetric extension of a Gauss-Bonnet like term is required in order to restore the supersymmetry invariance of the full Lagrangian.

Recently, in \cite{Hidden}, the authors clarified the role of the nilpotent fermionic generator $Q'$ introduced in \cite{D'AuriaFre} by looking at the gauge properties of the theory \footnote{They considered the $D=11$ and $D=7$ theories. In $D=11$, at least one extra spinor is required; In $D=7$, they found that two extra spinors are required. In this case, they also found two ``Lagrangian subalgebras", each one requiring one extra spinor.}.
They found that its presence is necessary, in order that the extra $1$-forms of the hidden superalgebra give rise to the correct gauge transformations of the $p$-forms of the FDA.
In particular, in its absence, the extra bosonic $1$-forms do \textit{not} enjoy gauge freedom, but generate, together with the supervielbein, new directions of an enlarged superspace, so that the FDA on ordinary superspace is no more reproduced.

On the group theoretical side, in \cite{Izaurieta} the authors developed the so-called \textit{$S$-expansion} procedure, which is based on combining the inner multiplication law of a discrete set $S$, which presents the structure of a semigroup, with the structure constants of a Lie algebra $\mathfrak{g}$. The new, larger Lie algebra thus obtained is called \textit{$S$-expanded algebra} and it is written as $\mathfrak{g}_S= S \times \mathfrak{g}$.

There are two facets applicable in the $S$-expansion method, which offer great manipulations on (super)algebras, \textit{i.e.} \textit{resonance} and \textit{reduction}. The role of \textit{resonance} is that of transferring the structure of the semigroup to the target (super)algebra; Meanwhile, \textit{reduction} plays a peculiar role in cutting the (super)algebra properly, thanks to the existence of a zero element in the set involved in the procedure.

From the physical point of view, several (super)gravity theories have been largely studied using the $S$-expansion approach, enabling the achievement of several results over recent years (see Ref.s \cite{Iza4, GRCS, CPRS1, Topgrav, BDgrav, Pat2,CR2, CRSnew, Static, Gen, Ein, Fierro1, Fierro2, Knodrashuk, Artebani, Concha1, Diego, Salgado, Concha2, Caroca:2010ax, Caroca:2010kr, Caroca:2011zz, Andrianopoli:2013ooa, Concha:2016hbt, Concha:2016kdz, Concha:2016tms, Durka:2016eun}).
Furthermore, in \cite{Marcelo2} an analytic method for $S$-expansion was developed. This method gives the multiplication table(s) of the (abelian) set(s) involved in an $S$-expansion process for reaching a target Lie (super)algebra from a starting one, after having properly chosen the partitions over subspaces of the considered (super)algebras.
A complete review of $S$-expansion can be found in \cite{Izaurieta} and \cite{Marcelo2}.

Recently, in \cite{PenRavera} the authors proposed a new prescription for $S$-expansion, involving
an infinite abelian semigroup $S^{(\infty)}$, with subsequent subtraction of a suitable infinite ideal.
Their approach is a generalization of the finite $S$-expansion procedure, and it allows to reproduce a generalized In\"on\"u-Wigner contraction (IW contraction) via infinite $S$-expansion between two different algebras.
Furthermore, the authors of \cite{lastpat} recently presented a generalization of the standard In\"on\"u-Wigner contraction, by rescaling not only the generators of a Lie superalgebra, but also the arbitrary constants appearing in the components of the invariant tensor.

In this work, we obtain a particular hidden Maxwell superalgebra in four dimensions by performing an infinite $S$-expansion with subsequent ideal subtraction of the hidden $AdS$-Lorentz superalgebra underlying $D=4$ supergravity.

The Maxwell (super)algebra corresponds to a
modification of the Poincar\'{e} (super)algebra.
In particular, in $D = 4$ the Maxwell superalgebra is obtained by adding to the Poincar\'{e} generators $\lbrace{J_{ab}, P_a\rbrace}$ the rank two tensorial charges $Z_{ab}$ \cite{bacry, schrader, soroka, gomis, bonanos, gibbons} \footnote{That is to say, by performing a tensor extension of the Poincar\'{e} algebra, as shown in \cite{soroka}. Furthermore, in the literature, $Z_{ab}$ has been associated with the presence of a constant electromagnetic background \cite{bacry, schrader, gomis, bonanos, gibbons}.}, and modifying the commutativity of the translation generators $P_a$ as follows:
\begin{equation}
[P_a, P_b] = Z_{ab}.
\end{equation}
In this way, the Maxwell algebra is an enlargement of the Poincar\'{e} algebra; Under the contraction $Z_{ab} \rightarrow 0$ we consistently recover the Poincar\'{e} algebra.
The generators $Z_{ab}$ commute with $P_a$, while having appropriate commutators with the Lorentz generators $J_{ab}$. We call the generators that commute with all the superalgebra but the Lorentz generators ``almost central".

In the literature, it was shown that the Maxwell algebra can be obtained as an
expansion procedure of the $AdS$ Lie algebra $so(3, 2)$ \cite{azcarraga, Salgado}. In particular, in Ref. \cite{Salgado} it was shown that the Maxwell algebra can be derived using the $S$-expansion procedure introduced in Ref. \cite{Izaurieta}. This result was then extended and generalized in \cite{Concha1}.
In Ref. \cite{Lukierski}, Maxwell superalgebras were discussed in the context of (generalized) Inon\"{u}-Wigner contraction, and in \cite{Lukierski2}, it was shown that the
$\mathcal{N} =1$, $D = 4$ Maxwell superalgebra can be obtained as an enlargement of the Poincar\'{e} superalgebra. This
is of particular interest, since it describes the supersymmetries of generalized $\mathcal{N} = 1$, $D = 4$ superspace in the presence of a constant abelian supersymmetric field strength background. Recently, it was shown that minimal Maxwell
superalgebra can be obtained using the Maurer-Cartan expansion method
\cite{azcarraga}. Subsequently, this superalgebra and its generalizations have been
obtained as $S$-expansions of the $AdS$ superalgebra \cite{Concha2}. This family
of superalgebras, which contain the Maxwell algebras type as bosonic
subalgebras, may be viewed as a generalization of the D'Auria-Fr\'{e} superalgebra introduced in \cite{D'AuriaFre} and Green algebras \cite{Green}.
Furthermore, in Ref. \cite{CR2}, the authors constructed the minimal $D = 4$ supergravity
action from the minimal Maxwell superalgebra, showing that $\mathcal{N} = 1$, $D = 4$ pure supergravity can be derived alternatively as a MacDowell-Mansouri like action from the minimal Maxwell superalgebra.

In the present work, we expand the hidden $AdS$-Lorentz superalgebra underlying $D=4$ supergravity, reaching a (hidden) Maxwell superalgebra. The latter can be viewed as an extension involving cosmological constant of the superalgebra underlying $D=4$ supergravity in flat spacetime.
The main reason for choosing the Maxwell superalgebra in four dimensions as a starting point of our analysis is that the almost central charges $Z_{ab}$ appearing in the Maxwell superalgebra, which can be discussed in terms of their dual description in the Maurer-Cartan formalism \cite{D'AuriaFre}, can be related to a bosonic $1$-form $B^{ab}$; The latter can be associated with an antisymmetric $3$-form $A^{(3)}$ on superspace, appropriately introduced in the context of Free Differential Algebras (see Ref.s \cite{D'AuriaFre, Castellani}, and \cite{Hidden} for further details), 
whose field strength is given by $F^{(4)}= dA^{(3)}$ (modulo gravitino $1$-form bilinears). 

The $3$-form $A^{(3)}$ does not give any dynamical contribution to the theory in four dimensions; In four-dimensional theories, $dA^{(3)}$ can be viewed as a (trivial) boundary term, while its field strength (which is proportional to the volume element in four dimensions) can be related to the presence of \textit{fluxes} (see Ref.s \cite{Vaula, Trigiante, Trigiante2,  Grana, Sam, Sam2, Bielleman}). In particular, in the context of string fluxes, in Ref. \cite{Bielleman} the authors discussed the role of Minkowski $3$-forms in flux string vacua, where all internal closed string fluxes are in one-to-one correspondence with quantized Minkowski $4$-forms. By performing a dimensional reduction of the $D=10$ Type II supergravity actions, they found that the $4$-forms act as auxiliary fields of the K\"{a}hler and complex structure moduli in effective actions. They also showed that the RR and NS axion dependence on the flux scalar potential appears through the said $4$-forms. Thus, they noticed that, although the Minkowski $3$-forms have no dynamical degrees of freedom in $D=4$, the kinetic terms of these $3$-forms lead to a Minkowski background which also contributes to the scalar potential of the theory. In addition, they also had some Chern-Simons couplings of the Minkowski $4$-forms to functions depending polinomially on the axionic fields and the internal fluxes.
 
In our case, we can say that the extra bosonic $1$-forms $B^{ab}$ and $\tilde{B}^{ab}$, which appear in the hidden Maxwell superalgebra (they are associated with the bosonic generator $Z_{ab}$ and $\tilde{Z}_{ab}$, respectively), in $D=4$ can be related to the presence of fluxes, since they are strictly related to the presence of the $3$-form $A^{(3)}$ in four dimensions (see Ref.s \cite{D'AuriaFre, Castellani, Hidden} for further details). In particular, the $1$-form $B^{ab}$ is directly associated with $A^{(3)}$, while the presence of $\tilde{B}^{ab}$ in the Maxwell superalgebra follows from the process of expansion considered in this paper. We do not analyze this aspect and the dynamics of the theory in the current work. Instead, we concentrate on the algebraic (hidden) structure of our model based on the hidden Maxwell superalgebra.

We adopt the FDAs the Maurer-Cartan formalism, and we write the parametrization of the $3$-form $A^{(3)}$, whose field strength is a $4$-form $F^{(4)}=dA^{(3)}+\ldots$, modulo fermionic bilinears, in terms of $1$-forms. We then show that the (trivial) boundary contribution in four dimensions, $dA^{(3)}$, can be naturally extended by considering particular contributions to the structure of the extra fermionic generator appearing in the hidden Maxwell superalgebra underlying supergravity in four dimensions. Its extensions involve the cosmological constant.
Interestingly, the presence of these terms strictly depends on the form of the extra fermionic generator appearing in the hidden superMaxwell-like extension of $D=4$ supergravity.

This paper is organized as follows: In Section \ref{ExpMax}, we perform expansions and contractions of different superalgebras describing and underlying $D=4$ supergravity, and we also display a map which links different superalgebras in four dimensions. In particular, we get a hidden Maxwell superalgebra in four dimensions by performing an infinite $S$-expansion with subsequent ideal subtraction of the hidden $AdS$-Lorentz superalgebra underlying $D=4$ supergravity.

In Section \ref{Main}, we write some of the superalgebras presented in Section \ref{ExpMax} in the Maurer-Cartan formalism and, in particular, we consider a hidden extension, involving cosmological constant, of $D=4$ supergravity, which corresponds to the (hidden) Maxwell superalgebra. We then write the parametrization of the $3$-form $A^{(3)}$ in this context and we show that the (trivial) boundary contribution $dA^{(3)}$ can be naturally extended with the addition of terms involving the cosmological constant. We also make some comment on the gauge invariance of our model.

Section \ref{Comments} contains the outlook and some possible future developments.
In the Appendix, we give detailed calculations on the infinite $S$-expansion with ideal subtraction performed on the hidden $AdS$-Lorentz superalgebra for reaching the hidden Maxwell superalgebra in four dimensions.

\section{Expansions and contractions of superalgebras in four dimensions}\label{ExpMax}

It is well known that we can construct several theories in four dimensions by choosing different amount of physical (and unphysical) fields, invariant under different symmetries. We can thus write different algebras and superalgebras. One of the simplest supersymmetric case is the Poincar\'e superalgebra $\overline{osp(1|4)}$, which is abelian in the momenta:
\begin{equation}
[P_a , P_b] = 0.
\end{equation}
On the other side, the (Anti-)de Sitter ($(A)dS$) algebra is characterized by the following commutation relation between translations:
\begin{equation}
[P_a,P_b]= J_{ab},
\end{equation}
where $J_{ab}$ are the Lorentz generators.
In Ref. \cite{McMan}, the authors presented a geometric formulation involving the $AdS$ structure group (the $AdS$ one), known as MacDowell-Mansouri action. The generalization of their work consists in considering the supergroup $OSp(\mathcal{N}|4)$.

In Ref. \cite{D'AuriaFre}, the authors presented a particular superalgebra, now known as \textit{hidden superalgebra}, underlying $D=11$ supergravity. This hidden superalgebra includes, as a subalgebra, the super-Poincar\'e algebra, and also involves two extra bosonic generators $Z_{ab}$ and $Z_{a_1 \ldots a_5}$ \footnote{Which were later understood as $p$-brane charges, sources of dual potentials $A^{(3)}$ and $B^{(6)}$, respectively (see Ref. \cite{townsend}).}, commuting with the generators $P_a$. Furthermore, they shown that an extra nilpotent fermionic generator $Q'$ must be included (in order to satisfy the closure of the superalgebra).

It is then possible to consider a \textit{hidden AdS-Lorentz superalgebra}, namely an extension of the $AdS$ superalgebra in which the commutators between the momenta is equal to a Lorentz-like generator, which will be referred as to $Z_{ab}$.
Finally, we should mentioned that the introduction of a second fermionic generator has been considered in the literature; Following this idea, the authors of \cite{Concha2} considered \textit{Maxwell superalgebras} for constructing actions for supergravity theories.

We will now consider (hidden) superalgebras in \textit{four dimensions}. Each of these superalgebras gives rise to the construction of an action for a supergravity theory. The existence of connections between different physical theories motivates to look for connections between the superalgebras underlying these theories.
We first consider a ``toy model" superalgebra in four dimensions described in \cite{Castellani}, namely an $AdS$-Lorentz-like superalgebra with an extra fermionic generator, which is the hidden superalgebra underlying the $AdS$ supergravity theory in $D=4$. This algebra will be named \textit{hidden $AdS$-Lorentz superalgebra}. It is generated by the set of generators $\lbrace{J_{ab},P_a,Z_{ab},Q_\alpha,Q'_\alpha \rbrace}$, and can be written as

\begin{align}
    \left[J_{ab},J_{cd}\right]=&\ \eta_{bc}J_{ad}-\eta_{ac}J_{bd}-\eta_{bd}J_{ac}+\eta_{ad}J_{bc}, & & \nonumber\\
\left[J_{ab},Z_{cd}\right]=&\ \eta_{bc}Z_{ad}-\eta_{ac}Z_{bd}-\eta_{bd}Z_{ac}+\eta_{ad}Z_{bc}, &&\nonumber\\
\left[Z_{ab},Z_{cd}\right]=&\ \eta_{bc}Z_{ad}-\eta_{ac}Z_{bd}-\eta_{bd}Z_{ac}+\eta_{ad}Z_{bc},\label{HiddenAdS}\\				
    \left[Q_\alpha ,Z_{ab}\right]=&-(\gamma_{ab}Q)_\alpha-(\gamma_{ab}Q')_\alpha , \;\;\; \left[Q'_\alpha ,Z_{ab}\right]=0, &\left[J_{ab},P_{c}\right]=&\ \eta_{bc}P_{a}-\eta_{ac}P_{b},\nonumber\\
\left[Q_\alpha ,P_a\right]=&-i (\gamma_a Q)_\alpha-i (\gamma_a Q')_\alpha ,  \;\;\; \left[Q'_\alpha ,P_a\right]=0, & \left[P_a,P_b\right]=&-Z_{ab},\nonumber\\
\left[J_{ab},Q_\alpha \right]=&-(\gamma_{ab}Q)_\alpha , 	\;\;\; \left[J_{ab},Q'_\alpha \right]=-(\gamma_{ab}Q')_\alpha ,  			&\left[Z_{ab},P_c\right]=&\ \eta_{bc}P_{a}-\eta_{ac}P_{b},\nonumber\\
\left\{Q_\alpha,Q_\beta \right\}=&-i(\gamma^a C)_{\alpha \beta}P_a-\frac{1}{2}(\gamma^{ab}C)_{\alpha \beta}Z_{ab},  &\left\{Q_\alpha,Q'_\beta \right\}= \left\{Q'_\alpha,Q'_\beta \right\} &=0, \nonumber
\end{align}
where $C$ stands for the charge conjugation matrix and $\gamma_a$ and $\gamma_{ab}$ are Dirac matrices in four dimensions. Let us notice
that the Lorentz type algebra generated by $\lbrace{J_{ab},Z_{ab} \rbrace}$ is a subalgebra of the above superalgebra.

In \cite{Gauss}, the authors explored the supersymmetry invariance of an extension of minimal $D=4$ supergravity in the presence of a non-trivial boundary, and they presented the explicit construction of
the $\mathcal{N}=1$, $D=4$ $AdS$-Lorentz supergravity bulk Lagragian in the rheonomic framework. In particular, they developed a peculiar way to
introduce a generalized supersymmetric cosmological term in supergravity.
The starting superalgebra they considered was a truncation of the hidden $AdS$-Lorentz one (\ref{HiddenAdS}). In fact, by performing a consistent truncation of the fermionic generator $Q'_\alpha$ in (\ref{HiddenAdS}), we get the $AdS$-Lorentz superalgebra considered in \cite{Gauss} \footnote{Let us observe that the authors of \cite{Gauss} adopted the Maurer-Cartan formalism in their work, where the superalgebra generators are properly associated to $1$-forms.}. In other words, the hidden $AdS$-Lorentz superalgebra can be consistently viewed as an extension of the $AdS$-Lorentz algebra described in \cite{Gauss}, with the inclusion of an extra fermionic generator $Q'_\alpha$.

On the other hand, the technique proposed by the authors of \cite{PenRavera}, which consists in a new prescription for $S$-expansion, involving an infinite abelian semigroup $S^{(\infty)}$, with subsequent subtraction of a suitable infinite ideal \footnote{Their approach is a generalization of the finite $S$-expansion procedure, and it  allows to reproduce a generalized In\"on\"u-Wigner contraction with an infinite $S$-expansion with subsequent ideal subtraction.}, allows to obtain a \textit{(hidden) Maxwell superalgebra} \cite{Lukierski, Lukierski2} in four dimensions, generated by the set of generators $\lbrace{J_{ab},P_a,Z_{ab},\tilde{Z}_{ab},Q_\alpha, \Sigma_\alpha \rbrace}$ (here and in the following, $\Sigma_\alpha$ denotes the extra nilpotent fermionic generator appearing in the hidden Maxwell superalgebra), by starting from the hidden $AdS$-Lorentz superalgebra (\ref{HiddenAdS}). Thus, following the approach described in \cite{PenRavera}, we can perform a $S$-expansion with the infinite abelian semigroup $S^{(\infty)}$ \footnote{The semigroup $S^{(\infty)}$ is an extension and generalization of the semigroups of the type $S^{(N)}_E = \lbrace{ \lambda_\alpha\rbrace}_{\alpha=0}^{N+1}$, endowed with the following multiplication rules: $\lambda_\alpha \lambda_\beta = \lambda_{\alpha+\beta}$ if $\alpha+\beta \leq N+1$, and $\lambda_\alpha \lambda_\beta = \lambda_{N+1}$ if $\alpha+\beta > N+1$.}, involving a resonant structure.
For further details on this calculation, see Appendix \ref{AppExp}. 

The hidden Maxwell superalgebra thus obtained reads
\begin{align}
    \left[J_{ab},J_{cd}\right]=&\ \eta_{bc}J_{ad}-\eta_{ac}J_{bd}-\eta_{bd}J_{ac}+\eta_{ad}J_{bc}, & & \nonumber\\
\left[J_{ab},Z_{cd}\right]=&\ \eta_{bc}Z_{ad}-\eta_{ac}Z_{bd}-\eta_{bd}Z_{ac}+\eta_{ad}Z_{bc}, &&\nonumber\\
\left[J_{ab},\tilde{Z}_{cd}\right]=&\ \eta_{bc}\tilde{Z}_{ad}-\eta_{ac}\tilde{Z}_{bd}-\eta_{bd}\tilde{Z}_{ac}+\eta_{ad}\tilde{Z}_{bc}, &&\nonumber\\
\left[Z_{ab},Z_{cd}\right]=&\ \eta_{bc}\tilde{Z}_{ad}-\eta_{ac}\tilde{Z}_{bd}-\eta_{bd}\tilde{Z}_{ac}+\eta_{ad}\tilde{Z}_{bc},\label{HiddensuperMaxwell}\\				
    \left[Q_\alpha ,Z_{ab}\right]=& -(\gamma_{ab}\Sigma)_\alpha , \;\;\; \left[\Sigma_\alpha ,Z_{ab}\right]=0, &\left[J_{ab},P_{c}\right]=&\ \eta_{bc}P_{a}-\eta_{ac}P_{b},\nonumber\\
\left[Q_\alpha ,P_a\right]=& -i (\gamma_a \Sigma)_\alpha ,  \;\;\; \left[\Sigma_\alpha ,P_a\right]=0, & \left[P_a,P_b\right]=&-\tilde{Z}_{ab},\nonumber\\
\left[J_{ab},Q_\alpha \right]=&-(\gamma_{ab}Q)_\alpha , 	\;\;\; \left[J_{ab},\Sigma_\alpha \right]=-(\gamma_{ab}\Sigma)_\alpha ,  			&\left[Z_{ab},P_c\right]=&0,\nonumber\\
\left\{Q_\alpha,Q_\beta \right\}=&-i(\gamma^a C)_{\alpha \beta}P_a-\frac{1}{2}(\gamma^{ab}C)_{\alpha \beta}Z_{ab},   & \left\{\Sigma_\alpha,\Sigma_\beta \right\} &=0, \nonumber\\
\left\{Q_\alpha,\Sigma_\beta \right\}=&-2 (\gamma^{ab}C)_{\alpha \beta}\tilde{Z}_{ab}, & &\nonumber \\
 \left[\tilde{Z}_{ab},P_{c}\right]=&\left[Q_\alpha,\tilde{Z}_{ab}\right]=\left[\Sigma_\alpha,\tilde{Z}_{ab}\right]=\left[\tilde{Z}_{ab},\tilde{Z}_{cd}\right]=0. & & \nonumber
\end{align}

It is well known that the Poincar\'e and the $AdS$ superalgebras are related by In\"on\"u-Wigner contraction, \textit{i.e.} by rescaling and consequently considering a particular limit for the generators. In the same way, by performing an In\"on\"u-Wigner contraction on the hidden $AdS$-Lorentz algebra (\ref{HiddenAdS}), we obtain the hidden Poincar\'e superalgebra (introduced and studied in \cite{D'AuriaFre, Castellani}). This superalgebra is generated by $\lbrace{J_{ab},P_a,Z_{ab},Q_\alpha,Q'_\alpha \rbrace}$, and can be written as

\begin{align}
\left[J_{ab},J_{cd}\right]=&\ \eta_{bc}J_{ad}-\eta_{ac}J_{bd}-\eta_{bd}J_{ac}+\eta_{ad}J_{bc}, &&\nonumber\\
\left[J_{ab},Z_{cd}\right]=&\ \eta_{bc}Z_{ad}-\eta_{ac}Z_{bd}-\eta_{bd}Z_{ac}+\eta_{ad}Z_{bc}, &&\nonumber\\
\left[J_{ab},P_{c}\right]=&\ \eta_{bc}P_{a}-\eta_{ac}P_{b}, \;\;\; \left[Z_{ab},Z_{cd}\right]=0, 			&\left[P_a,P_b\right]=&0,\nonumber\\
\left[Q_\alpha,P_a\right]=&-i(\gamma_aQ')_\alpha , \;\;\; \left[Q'_\alpha,P_{a}\right]=0, 						&\left[J_{ab},Q_\alpha \right]=&-(\gamma_{ab}Q)_\alpha, \label{poincare} \\
 \left[Q_\alpha ,Z_{ab}\right]=&-(\gamma_{ab}Q')_\alpha, \;\;\; \left[Q'_\alpha,Z_{ab}\right]=0, 						&\left[J_{ab},Q'_\alpha\right]=&-(\gamma_{ab}Q')_\alpha,\nonumber\\
\left\{Q_\alpha ,Q_\beta\right\}=&-i(\gamma^a C)_{\alpha \beta}P_a-\frac{1}{2}(\gamma^{ab}C)_{\alpha \beta}Z_{ab}	,			  &\left\{Q_\alpha,Q'_\beta \right\}=\left\{Q'_\alpha,Q'_\beta \right\}=&0. \nonumber
\end{align}

Let us observe that, analogously to the case of the hidden $AdS$-Lorentz and $AdS$-Lorentz superalgebras in four dimensions, a consistent truncation of the nilpotent fermionic generator $Q'_\alpha$ allows to reproduce the Poincar\'e superalgebra starting from the hidden Poincar\'e superalgebra (\ref{poincare}). In other words, algebra we call ``hidden Poincar\'e superalgebra" in four dimensions is an extension with one extra fermionic generator of the Poincar\'e superalgebra.

\begin{figure}
\begin{picture}(400,130)
  \put(10,40){\vector(0,-1){30}} 
   \put(15,27){{IW contraction}}
   \put(140,90){Hidden Maxwell}
 \put(0,45){$AdS$}
 \put(140,45){Hidden $AdS$-Lorentz}
   \put(160,70){{$S^{(\infty)}\ominus\mathcal{I}$}}
 \put(155,55){\vector(0,1){30}} 
 \put(302,45){$AdS$-Lorentz}
 \put(250,48){\vector(1,0){50}} 
    \put(255,55){$Q'_\alpha \rightarrow 0$}
 \put(0,0){Poincar\'e}
 \put(70,10){$Q'_\alpha \rightarrow 0$}
 \put(135,3){\vector(-1,0){80}} 
\put(35,45){\vector(1,0){100}} 
 \put(50,50){$S$-expansion}
   \put(160,25){{IW contraction}}
 \put(140,0){Hidden Poincar\'e}
  \put(155,42){\vector(0,-1){30}} 
\end{picture}
\caption{Map between different superalgebras in four dimensions. Here, $S^{(\infty)}\ominus\mathcal{I}$ denotes an infinite $S$-expansion with subsequent ideal subtraction \cite{PenRavera}.}\label{figure}
\end{figure}
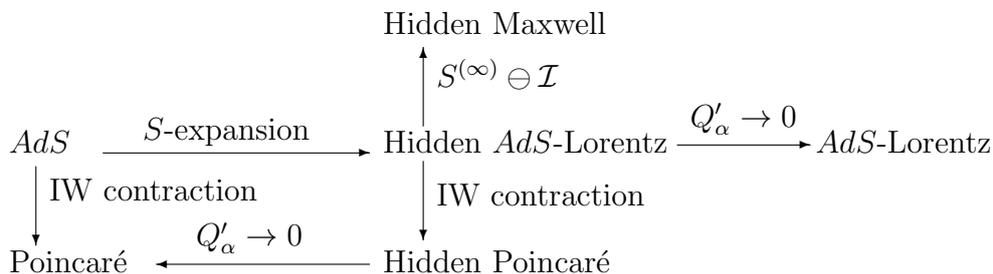

In Figure \ref{figure}, we have collected and summarized the relationships between the mentioned superalgebras \footnote{Let us remind that both the standard and the generalized In\"on\"u-Wigner contractions are reproducible through $S$-expansion: The standard In\"on\"u-Wigner contraction can be reproduced with a finite $S$-expansion, while the generalized one can be reproduced through an infinite $S$-expansions with subsequent ideal subtraction \cite{PenRavera}.}.

\section{Hidden Maxwell superalgebra in the Maurer-Cartan formalism and parametrization of the $3$-form $A^{(3)}$}\label{Main}

There are two dual ways of describing a (super)algebra: The first one is provided by the commutation relations between the generators; The second one is instead provided by the so-called Maurer-Cartan equations. These two descriptions are equivalent and dual each other.

The generators $T_A$'s, which form a basis of the tangent space $T(\mathcal{M})$ of a manifold $\mathcal{M}$, satisfy the commutation relations of the (super)algebra and the (super) Jacobi identity.
The same information is enclosed in the Maurer-Cartan equations, which read

\begin{equation}
d\sigma^A = -\frac{1}{2}C^A_{\ BC}\sigma^B \wedge \sigma^C,
\end{equation}
where $\sigma^A$ stands for the forms involved into the Maurer-Cartan equations, and where $C^A_{\ BC}$ are the structure constants. In the following, for simplicity, we will omit the symbol $\wedge$ denoting the wedge-product between forms.

As we can see, the Maurer-Cartan equations are written in terms of the dual $1$-forms $\sigma^A$'s of the generators $T_A$'s, which are related through the expression

\begin{equation}
 \sigma^A(T_B)=\delta^A_B,
\end{equation}
up to normalization factors \footnote{See the maps between the two formalism presented in \cite{Hidden} for further details.}.

We now consider the Maurer-Cartan equations associated with the superalgebras in $D=4$ presented in \cite{Castellani}. In the case of $\overline{osp(4 |1) }$ (Poincar\'e superalgebra), we have

\begin{align}
 R^{ab}=&0,\nonumber\\
 DV^a=&\frac{i}{2}\bar\Psi\gamma^a\Psi, \label{Mink}\\
 D\Psi=&0,\nonumber
\end{align}
where $\gamma^a$, as said before, are the four-dimensional gamma matrices, and where $D=d+\omega $ is the Lorentz covariant exterior derivative. Here we have fixed the normalization of $DV^a$ to $\frac{i}{2}$, according to the usual convention. The closure ($d^2=0$) of this superalgebra is trivially satisfied.

In the $AdS$ case, instead, the anticommutator of the fermionic generators $Q$'s falls into the Poincar\'e translations and the Lorentz rotations, generating non-vanishing value of the Lorentz curvature, namely

\begin{align}
R^{ab}=&\alpha e^2	 V^aV^b+\beta e\bar\Psi\gamma^{ab}\Psi,\nonumber\\
DV^a=&\frac{i}{2}\bar\Psi\gamma^a\Psi,\\
D\Psi=& i e\gamma_a\Psi V^a,\nonumber
\end{align}
where $e=1/2l$ corresponds to the inverse of the $AdS$ radius. Here, $\alpha$ and $\beta$ are parameters, and we have fixed the normalization of $D\Psi$ to $1$. From the closure requirement of the superalgebra ($d^2=0$), we get $\beta = \frac{1}{2}\alpha$ and, after having fixed the normalization $\alpha=-1$, we can thus write $\beta=-\frac{1}{2}$.

We observe that in the limit $e \rightarrow 0$ we correctly get the Maurer-Cartan equations in the flat (\textit{i.e.} Minkowski) space, namely equations (\ref{Mink}).

As shown in \cite{D'AuriaFre}, with the introduction of a nilpotent fermionic generator $Q'$ we can write the hidden Poincar\'e (\ref{poincare}) and the hidden $AdS$-Lorentz (\ref{HiddenAdS}) superalgebras in terms of the corresponding respective Maurer-Cartan equations. For the hidden Poincar\'e case (\ref{poincare}), we have

\begin{align}
 R^{ab}=&0,\\
 DV^a=&\frac{i}{2}\bar\Psi\gamma^a\Psi,\\
 D\Psi=&0,\\
 DB^{ab}=&\frac{1}{2}\bar\Psi\gamma^{ab}\Psi,\\
 D\eta=&\frac{i}{2}\delta\gamma_a\Psi V^a+\frac{1}{2}\epsilon\gamma_{ab}\Psi B^{ab}.
\end{align}
Here, $\delta$ and $\epsilon$ are two \textit{arbitrary} parameters. In fact, requiring the closure of the superalgebra, and in particular of $D\eta$, we simply get the identity $0=0$, which leads the solution to be given in terms of two free parameters, namely $\delta$ and $\epsilon$.
In particular, for reaching this result we have used the following Fierz identities in four dimensions:

\begin{equation}
\Psi \gamma_a \bar{\Psi}\gamma^a \Psi =0,
\end{equation},
\begin{equation}
\Psi \gamma_{ab} \bar{\Psi}\gamma^{ab} \Psi =0.
\end{equation}
As we can see, in this superalgebra the Lorentz curvature is zero: $R^{ab}=0$. However, we have a non-trivial ``$AdS$-like" contribution \footnote{We call this contribution ``$AdS$-like" since the $AdS$ curvature $\hat{R}^{ab}\equiv R^{ab} + e^2 V^a V^b+\frac{1}{2} e \bar{\Psi}\gamma^{ab}\Psi$ contains a similar term, namely the term involving the gravitino.} given by $DB^{ab}=\frac{1}{2}\bar\Psi\gamma^{ab}\Psi$.

Let us observe that in $D=4$ we also have a particular subalgebra of the hidden Poincar\'e one, which can be obtained through an In\"o\"u-Wigner contraction of the hidden $AdS$-Lorentz superalgebra (\ref{HiddenAdS}). In fact, we do not even need the $1$-form $B^{ab}$ to find an underlying group for the Cartan Integrable System (CIS) in the four dimensional Minkowski space. This subalgebra reads

\begin{align}
 R^{ab}=&0,\\
 DV^{a}=& \frac{i}{2}\bar\Psi\gamma^a\Psi,\\
 D\Psi=& 0,\\
 D\eta=& \frac{i}{2}\gamma_a\Psi V^a,
\end{align}
which endows the CIS with a $3$-form $A^{(3)}$ whose parametrization in terms of $1$-forms can be simply written as $A^{(3)}=-i \bar{\Psi}\gamma_a \eta V^a$ (see Ref. \cite{Castellani}).

As shown in Section \ref{ExpMax}, we can write a (hidden) Maxwell superalgebra in four dimensions, by starting from the hidden $AdS$-Lorentz one (\ref{HiddenAdS}). For completeness, in the following we report the Maurer-Cartan equations associated with the hidden $AdS$-Lorentz superalgebra (\ref{HiddenAdS}):
\begin{align}
 R^{ab}=&0,\\
 DV^a=&\frac{i}{2}\bar\Psi\gamma^a\Psi-e B^{ab}V_b,\\
 D\Psi=& \frac{i}{2}e \gamma_a\Psi V^a+\frac{e}{4}\gamma_{ab}\Psi B^{ab},\\
 DB^{ab}=&\frac{1}{2}\bar\Psi\gamma^{ab}\Psi - e B^{ac}B_{c}^{\; b}+e V^a V^b,\\
 D\eta=&\frac{i}{2}\gamma_a\Psi V^a+\frac{1}{4} \gamma_{ab}\Psi B^{ab},
\end{align}
where the parameters have been fixed by requiring the closure of the superalgbera and properly fixing the normalization of the $1$-form $\eta$ (see Ref. \cite{Castellani} for further details) \footnote{The authors of \cite{Castellani} observed that in the hidden AdS-Lorentz superalgebra we can write $D \eta=\frac{1}{e}\Lambda$ and $D \Psi = \Lambda$, where $\Lambda$ is the $2$-form that reads $\Lambda =\frac{i}{2}e\gamma_a \Psi V^a + \frac{1}{4}e \gamma_{ab}\Psi B^{ab}$.}.

We now write the Maurer-Cartan equations associated with the \textit{hidden Maxwell superalgebra} in $D=4$, namely

\begin{align}
 R^{ab}=&0,\\
 DV^a=&\frac{i}{2}\bar\Psi\gamma^a\Psi,\\
 D\Psi=&0,\\
 DB^{ab}=&\frac{1}{2}\bar\Psi\gamma^{ab}\Psi,\\
 D\tilde{B}^{ab}=&\alpha e\bar\Psi\gamma^{ab}\Phi + \beta e B^{ac}B_{c}^{\ b}+\gamma eV^aV^b,\\
 D\Phi = &\frac{i}{2}\delta\gamma_a\Psi V^a+\frac{1}{2}\epsilon \gamma_{ab}\Psi B^{ab},
 \end{align}
where $B^{ab}$ and $\tilde{B}^{ab}$ are the $1$-forms dual to the generators $Z_{ab}$ and $\tilde{Z}_{ab}$, respectively, and where $\Phi$ is the spinorial $1$-form dual to the extra nilpotent fermionic generator $\Sigma_\alpha$ appearing in the hidden Maxwell superalgebra.

Once again, we must require the closure $d^2=0$ of the superalgebra. In this way, from the first Maurer-Cartan equation we get $\delta \alpha = \gamma$, and $\beta = -2 \alpha \epsilon$.
We now choose the normalization $\alpha=1$ and $\delta=1$. We can thus write $\gamma=1$ and $\beta=-2\epsilon$, being $\epsilon$ a free parameter.
We observe that the Lorentz curvature is again zero: $R^{ab}=0$. In this case, we have two non-trivial ``$AdS$-like" contributions, namely $DB^{ab}=\frac{1}{2}\bar\Psi\gamma^{ab}\Psi$ and the term $\gamma e V^a V^b$ in $D\tilde{B}^{ab}=\alpha e\bar\Psi\gamma^{ab}\Phi+\beta e B^{ac}B_{c}^{\ b}+\gamma eV^aV^b$.
Then, we can finally write

\begin{align}
 R^{ab}=&0, \nonumber \\
 DV^a=&\frac{i}{2}\bar\Psi\gamma^a\Psi,  \nonumber\\
 D\Psi=&0 , \nonumber \\
 DB^{ab}=&\frac{1}{2}\bar\Psi\gamma^{ab}\Psi,  \nonumber\\
 D\tilde{B}^{ab}=& e\bar\Psi\gamma^{ab}\Phi+\beta e B^{ac}B_{c}^{\ b}+ eV^aV^b,  \nonumber\\
 D\Phi=&\frac{i}{2}\gamma_a\Psi V^a+\frac{1}{2}\epsilon \gamma_{ab}\Psi B^{ab}, \label{smalg}
\end{align}
where $\beta=-2\epsilon$.
This superalgebra is the \textit{hidden Maxwell superalgebra} underlying supergravity in four dimensions. 

We observe that setting $\beta=\epsilon=0$ we get the following subalgebra:

\begin{align}
 R^{ab}=&0,  \nonumber\\
 DV^a=&\frac{i}{2}\bar\Psi\gamma^a\Psi,  \nonumber \\
 D\Psi=&0,  \nonumber \\
 DB^{ab}=&\frac{1}{2}\bar\Psi\gamma^{ab}\Psi,  \nonumber \\
 D\tilde{B}^{ab}=& e\bar\Psi\gamma^{ab}\Phi + eV^aV^b,  \nonumber \\
 D\Phi=&\frac{i}{2}\gamma_a\Psi V^a. \label{smsubalg}
\end{align}

In the following, we will write the parametrization of the $3$-form $A^{(3)}$ appearing in the CIS in four-dimensional supergravity in terms of $1$-forms, both for the hidden Maxwell superalgebra (\ref{smalg}) and for its subalgebra (\ref{smsubalg}). We will then study the particular extensions of the (trivial) boundary contribution $dA^{(3)}$ in four dimensions.

\subsection{Extensions of $dA^{(3)}$ involving the cosmological constant}

We start our analysis by considering the hidden Maxwell superalgebra in four dimensions (\ref{smalg}). Then, we write the parametrization of the $3$-form $A^{(3)}$ in terms of $1$-forms, both for the hidden Maxwell superalgebra (\ref{smalg}) and for its subalgebra (\ref{smsubalg}), and we study the different extensions of $dA^{(3)}$. 

The $3$-form $A^{(3)}$,whose field strength in Minkowski space is given by $F^{(4)}= dA^{(3)}+ \frac{1}{2}\bar{\Psi}\gamma_{ab}\Psi V^a V^b$, does not give any dynamical contribution to the theory in four dimensions, and $dA^{(3)}$ can be viewed as a (trivial) boundary term. However, its field strength, which is proportional to the volume element in four dimensions, can be related to the presence of fluxes \cite{Vaula, Trigiante, Trigiante2, Grana, Sam, Sam2}. We can thus say that the extra bosonic $1$-forms, whose presence is related to the $3$-form $A^{(3)}$ (see \cite{D'AuriaFre, Castellani, Hidden} for further details) and which appear in the hidden Maxwell superalgebra in $D=4$, can be related to the presence of fluxes induced by $F^{(4)}$. In fact, $F^{(4)}$ can be written as
\begin{equation}
F^{(4)}\propto e \Omega,
\end{equation}
where $\Omega \propto \epsilon_{abcd} V^a V^b V^c V^d$ is the volume element in four dimensions; $F^{(4)}$ can thus be associated with a flux with charge $e$, being $e$ a constant parameter.
We do not develop this topic in our work and we do not analyze the dynamics of the theory. In the following, we concentrate on the pure \textit{algebraic structure} of the model.

Thus, let us now consider the hidden Maxwell superalgebra valued curvatures, which are defined by

\begin{align}
R^{ab}& \equiv d\omega ^{ab}-\omega ^a_{\; c}\omega ^{cb}\,, \\
R^{a}& \equiv DV^{a}-\frac{i}{2}\bar{\Psi }\gamma
^{a}\Psi \,, \\
F^{ab}& \equiv DB^{ab}-\frac{1}{2}\bar{\Psi}\gamma^{ab}\Psi \,, \\
\tilde{F}^{ab}& \equiv D\tilde{B}^{ab}- e \bar{\Psi}\gamma^{ab}\Phi -\beta e B^{ac}B_{c}^{\; b} -e V^a V^b\,, \\
\rho & \equiv D\Psi \,, \\
\zeta & \equiv D \Phi -\frac{i}{2}\gamma_a \Psi V^a -\frac{1}{2}\epsilon \gamma_{ab}\Psi B^{ab},
\end{align}
where $D=d+\omega $ is the Lorentz covariant exterior derivative.
In Minkowski space, we can also write

\begin{equation}
F^{(4)}  \equiv dA^{(3)}-\frac{1}{2}\bar{\Psi}\gamma_{ab}\Psi V^a V^b \,,
\end{equation}
where the $4$-form $F^{(4)}$ is trivially given in terms of a boundary contribution in four dimensions.
Our aim is that of writing the deformation to the $4$-form $F^{(4)}$ induced by the presence the cosmological constant in the hidden Maxwell superalgebra underlying $D=4$ supergravity.

We can write the Maurer-Cartan equations in four dimensions for the hidden Maxwell superalgebra, by setting the curvatures to zero in the vacuum, namely

\begin{align}
R^{ab}& \equiv d\omega ^{ab}-\omega ^a_{\; c}\omega ^{cb}=0\,, \\
R^{a}& \equiv DV^{a}-\frac{i}{2}\bar{\Psi }\gamma
^{a}\Psi=0 \,, \\
F^{ab}& \equiv DB^{ab}-\frac{1}{2}\bar{\Psi}\gamma^{ab}\Psi=0 \,, \\
\tilde{F}^{ab}& \equiv D\tilde{B}^{ab}- e \bar{\Psi}\gamma^{ab}\Phi -\beta e B^{ac}B_{c}^{\; b} -e V^a V^b=0\,, \\
\rho & \equiv D\Psi=0 \,, \\
\zeta & \equiv D \Phi -\frac{i}{2}\gamma_a \Psi V^a -\frac{1}{2}\epsilon \gamma_{ab}\Psi B^{ab}=0 \,,
\end{align}
which simply lead to the expression (\ref{smalg}).

Now, as done in the $D=11$ and $D=7$ supergravity cases in \cite{D'AuriaFre} and \cite{Hidden}, respectively, we can write the parametrization of the $3$-form $A^{(3)}$ in terms of $1$-forms. We first of all observe that, since $dA^{(3)}$ is a boundary contribution in four dimensions, we expect the topological form of $dA^{(3)}$ to lie in the parametrization of $A^{(3)}$.
We thus start by writing
\begin{equation}\label{a3parlast}
A^{(3)}=\frac{1}{2e}\bar{\Psi}\gamma_{ab}\Psi B^{ab}+\bar{\Psi}\gamma_{ab}\Phi \tilde{B}^{ab}+\beta \tilde{B}_{ab}B^{ac}B_{c}^{\;b}+\tilde{B}_{ab}V^aV^b-i\bar{\Psi}\gamma_a\Phi V^a-\epsilon \bar{\Psi}\gamma_{ab}\Phi B^{ab},
\end{equation}
where the topological structure is still \textit{not} manifest.
Let us observe that by setting $\beta = \epsilon=0$ in (\ref{a3parlast}) we obtain
\begin{equation}\label{a3bezero}
A^{(3)}=\frac{1}{2e}\bar{\Psi}\gamma_{ab}\Psi B^{ab}+\bar{\Psi}\gamma_{ab}\Phi \tilde{B}^{ab}+\tilde{B}_{ab}V^aV^b-i\bar{\Psi}\gamma_a\Phi V^a.
\end{equation}
Let us now reorganize and rewrite expression (\ref{a3parlast}) as follows:
\begin{equation}\label{paramA3}
A^{(3)}= \frac{1}{e}B^{ab}DB_{ab}+\frac{1}{e}\tilde{B}^{ab}D\tilde{B}_{ab}-2 \bar{\Phi}D\Phi .
\end{equation}
This particular parametrization will give rise to a topological structure for the boundary contribution $dA^{(3)}$. In fact, if we now consider the parametrization (\ref{paramA3}) and compute $dA^{(3)}$ , we get the following topological expression \footnote{The expression we get for $dA^{(3)}$ is said to be topological in the sense that the terms which appear in its structure are total derivatives, \textit{i.e.} boundary terms (trivial boundary terms in four dimensions).}:

\begin{align}
dA^{(3)}&= \frac{1}{e}d(B^{ab}DB_{ab})+\frac{1}{e}d(\tilde{B}^{ab}D\tilde{B}_{ab})-2 d(\bar{\Phi}D\Phi)= \nonumber \\
\label{da3}
&=\frac{1}{e}DB^{ab}DB_{ab}+\frac{1}{e}D\tilde{B}^{ab}D\tilde{B}_{ab}-2 D\bar{\Phi}D\Phi,
\end{align}
which automatically satisfies the closure requirement $d^2=0$.
If we now substitute the Maurer-Cartan equations (\ref{smalg}) in the expression (\ref{da3}), we get

\begin{align}
dA^{(3)}&=\frac{1}{2}\bar{\Psi}\gamma_{ab}\Psi V^a V^b+e\bar{\Psi}\gamma_{ab}\Phi \bar{\Psi}\gamma^{ab}\Phi +2\beta e \bar{\Psi}\gamma_{ab}\Phi B^{ac}B_{c}^{\;b}+2e \bar{\Psi}\gamma_{ab}\Phi V^a V^b+ \nonumber \\
\label{da3ext}
& +2\beta e B^{ac}B_c^{\;b}V_aV_b-2i \epsilon \bar{\Psi}\gamma_a \Psi B^{ab}V_b+\epsilon^2 \bar{\Psi}\gamma_{ac}\Psi B^{ab}B^c_{\; b}.
\end{align}
In the limit $e\rightarrow0$, the expression (\ref{da3ext}) reduces to

\begin{equation}\label{da3mc}
dA^{(3)}=\frac{1}{2}\bar{\Psi}\gamma_{ab}\Psi V^a V^b-2i \epsilon \bar{\Psi}\gamma_a \Psi B^{ab}V_b+\epsilon^2 \bar{\Psi}\gamma_{ac}\Psi B^{ab}B^c_{\; b}.
\end{equation}

We observe that, interestingly, this solution does \textit{not} reduce to the four-dimensional Minkowski flat space limit when $e\rightarrow0$. However, if we now consider the particular solution $\beta=\epsilon=0$, which conduces to the subalgebra (\ref{smsubalg}) of the hidden Maxwell superalgebra in four dimensions, we clearly see that, interestingly, this particular solution leads to

\begin{equation}
dA^{(3)}= \frac{1}{2}\bar{\Psi}\gamma_{ab}\Psi V^a V^b+e\bar{\Psi}\gamma_{ab}\Phi \bar{\Psi}\gamma^{ab}\Phi +2e \bar{\Psi}\gamma_{ab}\Phi V^a V^b,
\end{equation}
which exactly reproduces the Minkowski FDA with

\begin{equation}
dA^{(3)}=\frac{1}{2}\bar{\Psi}\gamma_{ab}\Psi V^aV^b
\end{equation}
in the limit $e \rightarrow 0$.
Thus, the particular subalgebra (\ref{smsubalg}) of the Maxwell superalgebra (\ref{smalg}) underlying supergravity in four dimensions can be written as

\begin{align}
R^{ab}&=0\,, \nonumber \\
DV^{a}&=\frac{i}{2}\bar{\Psi }\gamma
^{a}\Psi\,, \nonumber \\
DB^{ab}&=\frac{1}{2}\bar{\Psi}\gamma^{ab}\Psi \,, \nonumber \\
\label{MCeqnew}
D\tilde{B}^{ab}&= e \bar{\Psi}\gamma^{ab}\Phi +e V^a V^b \,,\nonumber \\
D\Psi &=0 \,, \nonumber \\
D \Phi &=\frac{i}{2}\gamma_a \Psi V^a  \,, \nonumber \\
dA^{(3)}&= \frac{1}{2}\bar{\Psi}\gamma_{ab}\Psi V^a V^b+e\bar{\Psi}\gamma_{ab}\Phi \bar{\Psi}\gamma^{ab}\Phi +2e \bar{\Psi}\gamma_{ab}\Phi V^a V^b,
\end{align}
where, having set $\beta= \epsilon=0$ in (\ref{smalg}) and (\ref{da3mc}), we have erased the $B^{ab}$-contributions in $D\tilde{B}^{ab}$ and $D\Phi$.

The hidden superalgebra (\ref{MCeqnew}) underlying $D=4$ supergravity is an extension involving cosmological constant of the hidden superalgebra underlying Poincar\'e supergravity in four dimensions. In particular, the superalgebra (\ref{MCeqnew}) is a subalgebra of the hidden Maxwell superalgebra obtained by performing an infinite $S$-expansion with subsequent ideal subtraction on the hidden $AdS$-Lorentz superalgebra underlying $D=4$ supergravity. 

In the FDAs' framework, the parametrization of the $3$-form $A^{(3)}$ appearing in the four-dimensional FDA presents a topological structure, which reflects on the (trivial) boundary contribution $dA^{(3)}$ (which can also be related to the presence of fluxes in $D=4$, as we have previously mentioned), as we can see from (\ref{da3}) and (\ref{MCeqnew}). 

Furthermore, the last expression in (\ref{MCeqnew}) consistently reproduces the FDA in Minkowski space, and in particular $dA^{(3)}= \frac{1}{2}\bar{\Psi}\gamma_{ab}\Psi V^a V^b$, when $e \rightarrow 0$. This new model underlying $D=4$ supergravity can be considered for the construction of a Lagrangian and for the study of the dynamics of the theory.

For the sake of completeness, we also observe that the parametrization (\ref{a3parlast}) can be also rewritten in the following form:
\begin{equation}\label{lastlastpar}
A^{(3)}= \tilde{B}_{ab}V^aV^b+\beta \tilde{B}_{ab}B^{ac}B_{c}^{\;b}-i\bar{\Psi}\gamma_a\Phi V^a+\bar{\Psi}\gamma_{ab}\left[\left(\frac{1}{2e}\Psi-\epsilon \Phi \right) B^{ab}+\Phi \tilde{B}^{ab}\right],
\end{equation}
where we remind that $\beta = -2\epsilon$, which shows us that the parametrization we have considered in the present work is given in terms of $1$-forms structures that are pretty similar to the ones appearing in the (``standard") parametrization of $A^{(3)}$ adopted in the Minkowski $D=11$ case in \cite{D'AuriaFre}, and later in \cite{Castellani}.
For $\beta=\epsilon=0$, the parametrization (\ref{lastlastpar}) becomes
\begin{equation}
A^{(3)}= \tilde{B}_{ab}V^aV^b-i\bar{\Psi}\gamma_a\Phi V^a+\bar{\Psi}\gamma_{ab}\left[\frac{1}{2e}\Psi B^{ab}+\Phi \tilde{B}^{ab}\right].
\end{equation}

Let us finally make an observation concerning gauge invariance: The supersymmetric FDA considered in this section is left invariant under the gauge transformation
\begin{equation}\label{gaugea3triv}
\delta A^{(3)} = d \Lambda^{(2)},
\end{equation}
generated by the arbitrary form $\Lambda^{(2)}$; The gauge transformations of the extra bosonic $1$-forms are
\begin{eqnarray}\label{gauge1f}
\left\{\begin{tabular}{l}
         $\delta B^{ab}=d\Lambda^{ab} $,\\
         $\delta \tilde{B}^{ab}=d\tilde{\Lambda}^{ab} $,
       \end{tabular}\right.
\end{eqnarray}
$\Lambda^{ab}$ and $\tilde{\Lambda}^{ab}$ being arbitrary Lorentz-valued scalar functions.
In \cite{Hidden}, the authors shown that the extra nilpotent fermionic generators appearing in the hidden superalgebras discussed in their work play the peculiar role of cohomological generators, allowing a fiber bundle structure on the group manifold and realizing the gauge invariance of the theory. In fact, in their presence the $3$-form $A^{(3)}$ parametrized in terms of $1$-forms transforms properly.

In our framework, by performing a simple algebraic calculation, we can see that there is no way for trivializing the gauge transformation of $A^{(3)}$ when dealing with the hidden Maxwell superalgebra described in (\ref{smalg}). In fact, in this case, once $A^{(3)}$ has been written in terms of bosonic and fermionic $1$-forms, one cannot get expression (\ref{gaugea3triv}) by exploiting the gauge transformations of the extra spinor $\Phi$. However, when considering the hidden subalgebra (\ref{smsubalg}), which is the one that consistently reduces to the Minkowski case in the limit $e\rightarrow 0$, one can prove that the gauge transformation of $A^{(3)}$ parametrized in terms of $1$-forms is simply given by
\begin{equation}\label{pargaugetr}
\delta A^{(3)} = d\left(\frac{1}{2 e}\bar{\Psi}\gamma_{ab}\Psi \Lambda^{ab}+\bar{\Psi}\gamma_{ab}\Phi \tilde{\Lambda}^{ab}+\tilde{\Lambda}_{ab}V^a V^b \right) = d \Lambda^{(2)},
\end{equation}
where we have defined
\begin{equation}
\Lambda^{(2)} \equiv \frac{1}{2 e}\bar{\Psi}\gamma_{ab}\Psi \Lambda^{ab}+\bar{\Psi}\gamma_{ab}\Phi \tilde{\Lambda}^{ab}+\tilde{\Lambda}_{ab}V^a V^b .
\end{equation}
We can thus see that equation (\ref{pargaugetr}) is equivalent to the requirement (\ref{gaugea3triv}) on the gauge transformation of the $3$-form.
In this case, even if we have no contribution to the gauge transformations coming from the extra spinor $\Phi$ (as we can see in (\ref{smsubalg})), the gauge transformation of the $3$-form $A^{(3)}$ results to be trivialized and can thus be written in the form (\ref{gaugea3triv}). 
Thus, we can say that the physical role of the extra spinor $\Phi$ appearing in the hidden Maxwell superalgebra (and in its subalgebra) studied in our work is different from the one described in \cite{Hidden}: In our model, in fact, it is necessary for the closure of the FDA, but it is not needed when requiring gauge invariance and trivialization of the gauge transformation of $A^{(3)}$.

Let us also observe that the model we have constructed is defined in a superspace that is \textit{larger} than the ordinary one (whose basis is defined by the supervielbein $\lbrace{V^a,\Psi \rbrace}$), and in the FDA developed in this work the extra $1$-forms must be considered as physical fields in an enlarged superspace.

\section{Comments and possible developments} \label{Comments}

In the present work, we have obtained a hidden Maxwell superalgebra underlying supergravity in four dimensions, by performing an infinite $S$-expansion of the hidden $AdS$-Lorentz superalgebra underlying the same theory, with subsequent ideal subtraction.

We have then written the hidden Maxwell superalgebra in the Maurer-Cartan formalism. The extra $1$-forms appearing in our approach in the hidden Maxwell superalgebra (and in its subalgebra) allow the closure of these superalgebras in four dimensions in an enlarged superspace. We have subsequently considered the parametrization of the $3$-form $A^{(3)}$ in terms of $1$-forms, in order to show the way in which the (trivial) boundary contribution in four dimensions, $dA^{(3)}$, can be naturally extended by considering particular contributions to the structure of the extra fermionic generator appearing in the hidden Maxwell superalgebra. These extensions involve the cosmological constant and, interestingly, their structure strictly depends on the form of the extra fermionic generators appearing in this hidden extension of $D=4$ supergravity.

Very recently, in Ref. \cite{LagrSalgado}, in the context of the so called the Chern-Simons-Antoniadis-Savvidy (ChSAS) forms, the authors used the
methods of FDA decomposition in terms of $1$-forms in order to construct a four-dimensional ChSAS supergravity
action for the Maxwell superalgebra, and they then used the Extended Cartan Homotopy
Formula to find a method that allows the separation of the ChSAS action into bulk and boundary contributions, and permits the splitting of the bulk Lagrangian into pieces that reflect the particular
subspace structure of the gauge algebra.

We conjecture that our approach could be useful in that context for the development of a supersymmetric Lagrangian with boundary in four dimensions, since we argue that, in the presence of a boundary, the contributions coming from $dA^{(3)}$ could help to restore supersymmetry in the Lagrangian given by bulk plus boundary contributions \footnote{The procedure would be analogous to the one described in \cite{Gauss} and \cite{bdy}.}. In other words, we think that the contributions coming from $dA^{(3)}$ could be able to restore (completely or in part \footnote{In the sense that one might need other terms, besides the ones coming from $dA^{(3)}$.}) the supersymmetry invariance of the complete Lagrangian (bulk plus boundary). 
In this context, one could also analyze the differences between the role played by the $dA^{(3)}$ contributions coming from the hidden Maxwell superalgebra and the ones coming from the subalgebra which properly reduces to the Minkowski case in the limit $e \rightarrow 0$. It would thus be interesting to write the Lagrangian in four dimensions considering boundary terms, by looking at the structure of $dA^{(3)}$, and to study other possible extensions of $dA^{(3)}$ depending on the cosmological constant, when dealing with different (hidden) superalgebras. One may also construct a Lagrangian based on the hidden Maxwell superalgebra described in this work by considering the components of the invariant tensor of this target superalgebra, obtained through infinite $S$-expansion with ideal subtraction, written in terms of those of the starting algebra \cite{PenRavera,CR2}.

Furthermore, it would be interesting to analyze the relation between fluxes and gauge algebras in $D=4$ effective supergravities.

Another possible development of the present work would be the study of Chern-Simons theories and Lagrangians in even dimensions \cite{LagrSalgado}, such as the four dimensional case, in the context of (hidden) Maxwell superalgebras, and Born-Infeld theories, since they are topological theories and they can be affected by the presence of a non-trivial boundary. Furthermore, this study can also be extended to theories in higher dimensions and to higher spin cases, and one can construct topological supergravities by considering the transgression field theory scheme described in \cite{transgression}.

Finally, it would be interesting to consider the family of Maxwell superalgebras introduced in \cite{Concha:2016hbt} and to discuss them in the context of FDAs in four and higher dimensions, in order to be able to analyze the possible extensions of the higher degree forms in this framework, by following the approach presented in this work.

\section{Acknowledgements}

We are grateful to L. Andrianopoli for a critical reading of the manuscript and for the enlightening comments and suggestions, and to R. D'Auria and M. Trigiante for the insightful discussions and the constant support.

One of the authors (D. M. Pe\~{n}afiel) was supported by grants from the \textit{Comisi\'{o}n Nacional de Investigaci\'{o}n Cient\'{i}fica y Tecnol\'{o}gica} CONICYT and from the \textit{Universidad de Concepci\'{o}n}, Chile, and wishes to thank L. Andrianopoli, R. D'Auria, and M. Trigiante for their kind hospitality at DISAT - \textit{Dipartimento di Scienza Applicata e Tecnologia} of the \textit{Polytechnic of Turin}, Italy.

\appendix

\section{Infinite $S$-expansion with ideal subtraction of the hidden $AdS$-Lorentz superalgebra in four dimensions}\label{AppExp}

In the following, we adopt the technique proposed in \cite{PenRavera}, namely an infinite $S$-expansion involving an abelian semigroup $S^{(\infty)}$, with subsequent subtraction of a suitable ideal, in order to obtain the hidden Maxwell superalgebra in four dimensions (\ref{HiddensuperMaxwell}), generated by the set of generators $\lbrace{J_{ab},P_a,Z_{ab},\tilde{Z}_{ab},Q_\alpha, \Sigma_\alpha \rbrace}$, by starting from the hidden $AdS$-Lorentz superalgebra (\ref{HiddenAdS}), generated by $\lbrace{J_{ab},P_a,Z_{ab},Q_\alpha, Q'_\alpha \rbrace}$.

Thus, we consider the commutation relations of the hidden $AdS$-Lorentz superalgebra in four dimensions (\ref{HiddenAdS}). We report them here for completeness:

\begin{align}
    \left[J_{ab},J_{cd}\right]=&\ \eta_{bc}J_{ad}-\eta_{ac}J_{bd}-\eta_{bd}J_{ac}+\eta_{ad}J_{bc}, & & \nonumber\\
\left[J_{ab},Z_{cd}\right]=&\ \eta_{bc}Z_{ad}-\eta_{ac}Z_{bd}-\eta_{bd}Z_{ac}+\eta_{ad}Z_{bc}, &&\nonumber\\
\left[Z_{ab},Z_{cd}\right]=&\ \eta_{bc}Z_{ad}-\eta_{ac}Z_{bd}-\eta_{bd}Z_{ac}+\eta_{ad}Z_{bc},\label{HiddenAdSAPP}\\				
    \left[Q_\alpha ,Z_{ab}\right]=&-(\gamma_{ab}Q)_\alpha-(\gamma_{ab}Q')_\alpha , \;\;\; \left[Q'_\alpha ,Z_{ab}\right]=0, &\left[J_{ab},P_{c}\right]=&\ \eta_{bc}P_{a}-\eta_{ac}P_{b},\nonumber\\
\left[Q_\alpha ,P_a\right]=&-i (\gamma_a Q)_\alpha-i (\gamma_a Q')_\alpha ,  \;\;\; \left[Q'_\alpha ,P_a\right]=0, & \left[P_a,P_b\right]=&-Z_{ab},\nonumber\\
\left[J_{ab},Q_\alpha \right]=&-(\gamma_{ab}Q)_\alpha , 	\;\;\; \left[J_{ab},Q'_\alpha \right]=-(\gamma_{ab}Q')_\alpha ,  			&\left[Z_{ab},P_c\right]=&\ \eta_{bc}P_{a}-\eta_{ac}P_{b},\nonumber\\
\left\{Q_\alpha,Q_\beta \right\}=&-i(\gamma^a C)_{\alpha \beta}P_a-\frac{1}{2}(\gamma^{ab}C)_{\alpha \beta}Z_{ab},  &\left\{Q_\alpha,Q'_\beta \right\}= \left\{Q'_\alpha,Q'_\beta \right\} &=0. \nonumber
\end{align}

We first of all perform an In\"{o}n\"{u}-Wigner contraction on the generator $Q'$. 
Then, we consider the subspace partition $V_0= \lbrace{J_{ab} \rbrace}$, $V_1= \lbrace{Q_\alpha \rbrace}$, $V_2= \lbrace{Z_{ab} \rbrace}$, $V_3=  \lbrace{P_{a} \rbrace}$, and we perform an infinite $S$-expansion with the infinite semigroup $S^{(\infty)} = \lbrace{\lambda_0,\lambda_1,\lambda_2,\ldots, \infty \rbrace}$, namely
\begin{align}
\hat{V}_0 & = \lbrace{\lambda_0,\lambda_1,\lambda_2,\lambda_3,\lambda_4,\ldots, \infty \rbrace}\times \lbrace{ J_{ab}\rbrace}, \\
\hat{V}_1 & = \lbrace{\lambda_0,\lambda_1,\lambda_2,\lambda_3,\lambda_4,\ldots, \infty \rbrace}\times \lbrace{ Q_\alpha \rbrace}, \\
\hat{V}_2 & = \lbrace{\lambda_0,\lambda_1,\lambda_2,\lambda_3,\lambda_4,\ldots, \infty \rbrace}\times \lbrace{ Z_{ab}\rbrace}, \\
\hat{V}_3 & = \lbrace{\lambda_0,\lambda_1,\lambda_2,\lambda_3,\lambda_4,\ldots, \infty \rbrace}\times \lbrace{ Q'_\alpha\rbrace}, 
\end{align}
where $\hat{V}_i, \; i=0,1,2,3$ are the subspaces of the target superalgebra (here and in the following, we will refer to the quantities related to the target superalgebra as to quantities with the upper $\hat{}$ symbol).

Let us remind that the semigroup $S^{(\infty)}$ is an extension and generalization of the semigroups of the type $S^{(N)}_E = \lbrace{ \lambda_\alpha\rbrace}_{\alpha=0}^{N+1}$, endowed with the multiplication rules $\lambda_\alpha \lambda_\beta = \lambda_{\alpha+\beta}$ if $\alpha+\beta \leq N+1$, and $\lambda_\alpha \lambda_\beta = \lambda_{N+1}$ if $\alpha+\beta > N+1$.

Then we perform, by following the procedure described in \cite{PenRavera}, the subtraction of the infinite ideal given by
\begin{equation}
\mathcal{I}= W_0 \oplus W_1 \oplus W_2 \oplus W_3,
\end{equation}
where
\begin{align}
W_0 = & \lbrace{ \lambda_{2l}, \; l= 1, \ldots, \infty \rbrace} , \\
W_1 = & \lbrace{\lambda_{2m+1}, \; m= 2, \ldots, \infty \rbrace} , \\
W_2 = & \lbrace{ \lambda_{2n}, \; n= 3, \ldots, \infty \rbrace} , \\
W_3 = & \lbrace{ \lambda_{2p}, \; l= 2, \ldots, \infty \rbrace} .
\end{align}

We then define
\begin{align}
\hat{J}_{ab} &\equiv \lambda_0 J_{ab}, \\
\hat{Z}_{ab} &\equiv \lambda_2 Z_{ab}, \\
\hat{\tilde{Z}}_{ab} &\equiv \lambda_4 Z_{ab}, \\
\hat{Q}_{\alpha} &\equiv \lambda_1 Q_{\alpha}, \\
\hat{\Sigma}_{\alpha} &\equiv \lambda_3 Q_{\alpha}, \\
\hat{P}_{a} &\equiv \lambda_2 P_{a}. \\
\end{align}

If we now write the expansion of the commutation relations (\ref{HiddenAdSAPP}), and we rename the target generators by simply removing the upper $\hat{}$ symbol, namely
\begin{align}
\hat{J}_{ab}  & \rightarrow J_{ab}, \\
\hat{Z}_{ab}  & \rightarrow Z_{ab}, \\
\hat{\tilde{Z}}_{ab}  & \rightarrow \tilde{Z}_{ab}, \\
\hat{P}_{a}  & \rightarrow P_{a}, \\
\hat{Q}_{\alpha}  & \rightarrow Q_{\alpha}, \\
\hat{\Sigma}_{\alpha}  & \rightarrow \Sigma_{\alpha},
\end{align}
we finally end up with the hidden Maxwell superalgebra in four dimensions (\ref{HiddensuperMaxwell}), generated by $\lbrace{J_{ab},Z_{ab},\tilde{Z}_{ab},P_a,Q_\alpha,\Sigma_\alpha \rbrace}$. For the sake of completeness, we also report its commutation relations in the following:

\begin{align}
    \left[J_{ab},J_{cd}\right]=&\ \eta_{bc}J_{ad}-\eta_{ac}J_{bd}-\eta_{bd}J_{ac}+\eta_{ad}J_{bc}, & & \nonumber\\
\left[J_{ab},Z_{cd}\right]=&\ \eta_{bc}Z_{ad}-\eta_{ac}Z_{bd}-\eta_{bd}Z_{ac}+\eta_{ad}Z_{bc}, &&\nonumber\\
\left[J_{ab},\tilde{Z}_{cd}\right]=&\ \eta_{bc}\tilde{Z}_{ad}-\eta_{ac}\tilde{Z}_{bd}-\eta_{bd}\tilde{Z}_{ac}+\eta_{ad}\tilde{Z}_{bc}, &&\nonumber\\
\left[Z_{ab},Z_{cd}\right]=&\ \eta_{bc}\tilde{Z}_{ad}-\eta_{ac}\tilde{Z}_{bd}-\eta_{bd}\tilde{Z}_{ac}+\eta_{ad}\tilde{Z}_{bc},\label{HiddensuperMaxwellAPP}\\				
    \left[Q_\alpha ,Z_{ab}\right]=& -(\gamma_{ab}\Sigma)_\alpha , \;\;\; \left[\Sigma_\alpha ,Z_{ab}\right]=0, &\left[J_{ab},P_{c}\right]=&\ \eta_{bc}P_{a}-\eta_{ac}P_{b},\nonumber\\
\left[Q_\alpha ,P_a\right]=& -i (\gamma_a \Sigma)_\alpha ,  \;\;\; \left[\Sigma_\alpha ,P_a\right]=0, & \left[P_a,P_b\right]=&-\tilde{Z}_{ab},\nonumber\\
\left[J_{ab},Q_\alpha \right]=&-(\gamma_{ab}Q)_\alpha , 	\;\;\; \left[J_{ab},\Sigma_\alpha \right]=-(\gamma_{ab}\Sigma)_\alpha ,  			&\left[Z_{ab},P_c\right]=&0,\nonumber\\
\left\{Q_\alpha,Q_\beta \right\}=&-i(\gamma^a C)_{\alpha \beta}P_a-\frac{1}{2}(\gamma^{ab}C)_{\alpha \beta}Z_{ab},   & \left\{\Sigma_\alpha,\Sigma_\beta \right\} &=0, \nonumber\\
\left\{Q_\alpha,\Sigma_\beta \right\}=&-2 (\gamma^{ab}C)_{\alpha \beta}\tilde{Z}_{ab}, & &\nonumber \\
 \left[\tilde{Z}_{ab},P_{c}\right]=&\left[Q_\alpha,\tilde{Z}_{ab}\right]=\left[\Sigma_\alpha,\tilde{Z}_{ab}\right]=\left[\tilde{Z}_{ab},\tilde{Z}_{cd}\right]=\left[Z_{ab},\tilde{Z}_{cd}\right]=0. & & \nonumber
\end{align}

\end{document}